\documentclass[twocolumn,aps,prc,superscriptaddress,showpacs,floatfix]{revtex4-1}
\usepackage{url}
\usepackage{cancel}
\usepackage[colorlinks,linkcolor=blue,citecolor=blue,filecolor=black,urlcolor=blue]{hyperref}
\usepackage{epsfig,graphics}
\usepackage{graphicx}
\usepackage{dcolumn}
\usepackage{bm}
\usepackage[usenames]{color}
\usepackage{amssymb}
\usepackage{amsmath}
\usepackage{multirow}
\usepackage{float}
\usepackage{harpoon}
\usepackage{MnSymbol}
\usepackage{appendix}
\usepackage{color}
\usepackage{hyperref}
\usepackage{cleveref}

\newcommand{\sqrtsnn}{\mbox{$\sqrt{s^{}_{\mathrm{NN}}}$}}

\newcommand{\lr}[1]{\left\langle #1\right\rangle}

\newcommand{\auau}{$^{197}$Au+$^{197}$Au}
\newcommand{\ruru}{$^{96}$Ru+$^{96}$Ru}
\newcommand{\zrzr}{$^{96}$Zr+$^{96}$Zr}
\newcommand{\uu}{$^{238}$U+$^{238}$U}

\begin{document}
\title{Measuring deformed neutron skin with free spectator nucleons in relativistic heavy-ion collisions}
\author{Lu-Meng Liu}
\affiliation{School of Physical Sciences, University of Chinese Academy of Sciences, Beijing 100049, China}
\author{Jun Xu}\email[Correspond to\ ]{xujun@zjlab.org.cn}
\affiliation{School of Physics Science and Engineering, Tongji University, Shanghai 200092, China}
\affiliation{Shanghai Advanced Research Institute, Chinese Academy of Sciences, Shanghai 201210, China}
\affiliation{Shanghai Institute of Applied Physics, Chinese Academy of Sciences, Shanghai 201800, China}
\author{Guang-Xiong Peng}
\affiliation{School of Nuclear Science and Technology, University of Chinese Academy of Sciences, Beijing 100049, China}
\affiliation{Theoretical Physics Center for Science Facilities, Institute of High Energy Physics, Beijing 100049, China}
\affiliation{Synergetic Innovation Center for Quantum Effects $\&$ Applications, Hunan Normal University, Changsha 410081, China}
\date{\today}

\begin{abstract}
The neutron skin in deformed nuclei is generally not uniformly distributed but has an angular distribution, depending on both the spin-dependent nuclear interaction and the nuclear symmetry energy. To extract the information of the deformed neutron skin, we have explored the possibility of using free spectator nucleons in central tip-tip and body-body collisions at top RHIC energy with four typical deformed nuclei. The density distributions of neutrons and protons are consistently obtained from the Skyrme-Hartree-Fock-Bogolyubov calculation, and the angular distribution of the neutron skin can be varied by adjusting the strength of the nuclear spin-orbit coupling. With the information of spectator nucleons obtained based on a Monte-Carlo Glauber model, the free spectator nucleons are generated from a multifragmentation process. By investigating the results from different systems and with different collision configurations, we found that although it is difficult to probe the deformed neutron skin in $^{96}$Zr and $^{238}$U by their collisions, it is promising to extract the polar angular distributions of the neutron skin in $^{96}$Ru and $^{197}$Au by comparing the yield ratios of free spectator neutrons to protons in their central tip-tip and body-body collisions. The proposed observables can be measured by dedicated zero-degree calorimeters in heavy-ion collision experiments that have been carried out in recent years by RHIC.
\end{abstract}
\maketitle


Nucleon distribution inside a nucleus is a fundamental probe of the nuclear interaction and the nuclear matter equation of state (EOS). The neutron-skin thickness $\Delta r_{\mathrm{np}}$, i.e., generally defined as the difference between the neutron and proton root-mean-square (RMS) radii, is a robust probe of the slope parameter $L$ of the nuclear symmetry energy~\cite{Horowitz:2000xj,Furnstahl:2001un,Todd-Rutel:2005yzo,Centelles:2008vu,Zhang:2013wna,Xu:2020fdc}, characterizing the isospin dependence of the nuclear matter EOS. The density distributions in most nuclei, especially in the vicinity of full shell or subshell, are deformed, largely affected by the nuclear spin-orbit coupling (SOC)~\cite{book1,book2}. Naively, one expects that the neutron skin is also deformed in deformed nuclei, and may play a role in nucleus collective excitations, e.g., the oscillation of the neutron skin in a scissor like motion against the proton-neutron
core in deformed nuclei~\cite{Arteaga:2009in}. While early studies claimed that the neutron skin is roughly isotropic and independent of the polar angle in deformed nuclei~\cite{Hamamoto:1995zz,Sarriguren:2007pw}, we will show in the present study that its angular distribution depends on the particular nucleus and the strength of the nuclear SOC. As is known, the nuclear SOC is crucial for explaining successfully the magic numbers for stable nuclei~\cite{PhysRev.75.1969,PhysRev.75.1766.2}, and its strength affects the property of drip-line nuclei~\cite{LALAZISSIS19987}, the astrophysical r-process~\cite{CHEN199537}, and the location of the island of stability for superheavy elements~\cite{PhysRevC.60.034304,PhysRevLett.101.072701}. Measuring the deformed neutron skin can thus be helpful for understanding properties of the nuclear SOC and its interplay with the nuclear symmetry energy.

The $\Delta r_{\mathrm{np}}$ can be measured experimentally through proton~\cite{Zenihiro:2010zz,Terashima:2008rb} and pion~\cite{Friedman:2012pa} scatterings, charge exchange reactions~\cite{Krasznahorkay:1999zz}, coherent pion photoproductions~\cite{Tarbert:2013jze}, antiproton annihilations~\cite{Klos:2007is,Brown:2007zzc,Trzcinska:2001sy}, and parity-violating electron-nucleus scatterings~\cite{PhysRevLett.126.172502,CREX:2022kgg}. However, these traditional methods measure the average $\Delta r_{\mathrm{np}}$ and mostly in spherical nuclei. Relativistic heavy-ion collisions, in which the nucleon distribution determines the initial condition, provide a unique way of measuring both the neutron-skin thickness and the deformation of colliding nuclei~\cite{Filip:2009zz,Shou:2014eya,Giacalone:2019pca,Jia:2021tzt,Li:2019kkh,Jia:2021oyt,Xu:2021uar,Jia:2021qyu,Giacalone:2019pca,Bally:2021qys,Jia:2021wbq}. One expects that observables could be affected by the angular distribution of the $\Delta r_{\mathrm{np}}$ in deformed nuclei if typical collision configurations, such as tip-tip (with symmetric axis head-on) and body-body (head-on but with symmetric axis parallel) collisions, can be selected. Although it is very challenging to select events of special orientations in high-energy collisions with deformed nuclei, several promising triggers have been proposed in the literature (see, e.g., Refs.~\cite{Nepali:2007an,Goldschmidt:2015kpa}).

While the recent isobaric collisions, i.e., \ruru\ and \zrzr\ collisions at $\sqrtsnn=200$ GeV, are unable to detect considerable chiral magnetic effect~\cite{STAR:2021mii}, various observables at midrapidities were proposed as probes of the $\Delta r_{\mathrm{np}}$ in colliding nuclei~\cite{Li:2019kkh,Jia:2021oyt,Xu:2021uar,Jia:2021qyu,Giacalone:2019pca,Bally:2021qys,Jia:2021wbq}. Recently, we have proposed that the yield ratio of free spectator neutrons, which are measurable by zero-degree calorimeters, in ultracentral \zrzr\ to \ruru\ collision systems, can be a robust probe of the $\Delta r_{\mathrm{np}}$ in colliding nuclei~\cite{Liu:2022kvz}, free from the uncertainties of modeling the complicated dynamics in the midrapidity region. We have further proposed in Ref.~\cite{Liu:2022xlm} that the yield ratio $N_n/N_p$ of free spectator neutrons to protons in a single collision system can be a more sensitive probe of the $\Delta r_{\mathrm{np}}$, if spectator protons can also be measured by instrumenting the forward region with dedicated detectors~\cite{Tarafdar:2014oua}. In the previous studies, we consider \zrzr\ and \ruru\ collisions with random orientations. Since $^{96}$Zr and $^{96}$Ru are both deformed nuclei, e.g., both quadrupole ($\beta_2=0.06$) and octupole ($\beta_3=0.20$) deformation for $^{96}$Zr and a quadrupole deformation ($\beta_2=0.16$) for $^{96}$Ru are extracted in a recent flow analysis~\cite{Zhang:2021kxj}, it is promising to measure the angular distribution of $\Delta r_{\mathrm{np}}$ in $^{96}$Zr and $^{96}$Ru by selecting special collision configurations in the recent isobaric collisions. Besides $^{96}$Zr and $^{96}$Ru, $^{238}$U is a famous neutron-rich and prolate-shaped nucleus with a quadrupole deformation of about $\beta_2=0.28$~\cite{Moller:2015fba}, and \uu\ collisions at \sqrtsnn=193 GeV were carried out by the STAR Collaboration in recent years~\cite{STAR:2015mki,STAR:2021twy}. Moreover, a scaling analysis of the elliptic flow at RHIC energy from colliding nuclei with different quadrupole deformations shows that the widely used $^{197}$Au is an oblate-shaped nucleus with a quadrupole deformation of about $\beta_2=-0.15$~\cite{Giacalone:2021udy}. In the present study, by selecting special collision configurations, e.g., central tip-tip and body-body collisions, for \zrzr, \ruru, and \auau\ collisions at $\sqrtsnn=200$ GeV and \uu\ collisions at $\sqrtsnn=193$ GeV, we explore the possibility of measuring the polar angular distribution of the neutron skin in colliding nuclei.

\begin{figure}[!h]
\includegraphics[width=0.8\linewidth]{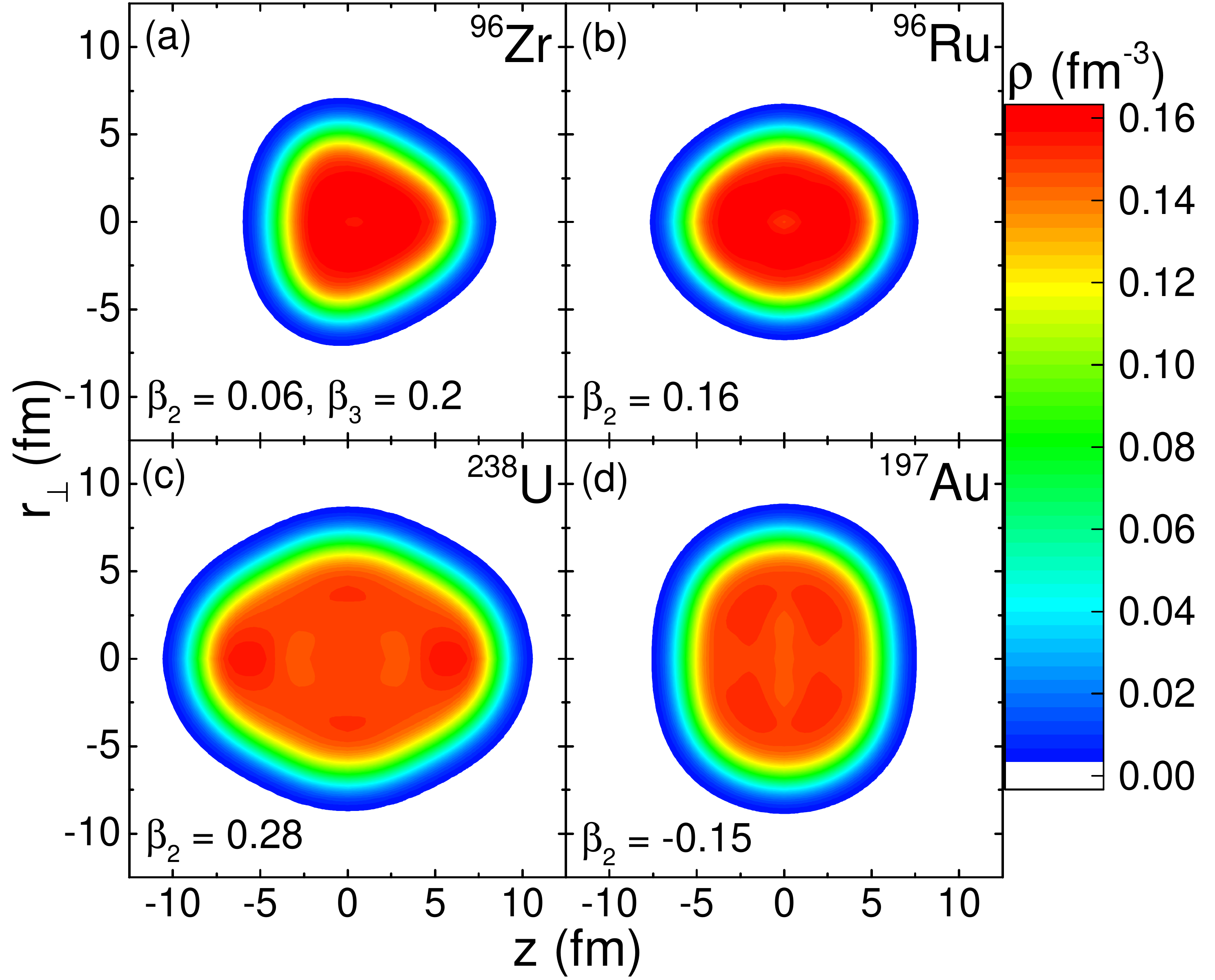}
\vspace*{-.3cm}
\caption{\label{fig1} Density contours of nucleons in the $r_\perp-z$ plane for $^{96}$Zr with $\beta_2=0.06$ and $\beta_3=0.2$ (a), $^{96}$Ru with $\beta_2=0.16$ (b), $^{238}$U with $\beta_2=0.28$ (c), and $^{197}$Au with $\beta_2=-0.15$ (d) from deformed SHFB calculations by using $L=90$ MeV and $W_0=133$ MeVfm$^5$.}
\end{figure}

We obtain the nucleon density distributions in deformed nuclei based on the Skyrme-Hartree-Fock-Bogolyubov (SHFB) calculation \cite{Stoitsov:2012ri}, where the energy-density functional originates from the standard effective Skyrme interaction~\cite{Chen:2010qx} through the Hartree-Fock method, among which the effective spin-orbit interaction between two nucleons at positions $\vec{r}_1$ and $\vec{r}_2$ can be expressed as~\cite{PhysRevC.5.626}
\begin{equation}
v_{so} = i W_0 (\vec{\sigma}_1+\vec{\sigma}_2) \cdot \vec{k}^\prime \times
\delta(\vec{r}_1-\vec{r}_2) \vec{k},
\end{equation}
where $W_0$ is the strength of the spin-orbit coupling whose default value is set to be 133 MeV fm$^5$ and is generally constrained within $W_0=80-150$ MeVfm$^5$ based on nuclear structure studies~\cite{PhysRevC.76.014312,PhysRevC.77.024316,PhysRevC.80.064302}, $\vec{\sigma}_{1(2)}$ represents the Pauli matrices, $\vec{k}=(\vec{p}_1-\vec{p}_2)/2$ is the relative momentum operator acting on the right with $\vec{p}=-i\nabla$, and $\vec{k}^\prime$ is the complex conjugate of $\vec{k}$. With the Hartree-Fock method, the above spin-orbit interaction leads to the potential energy density expressed as~\cite{PhysRevC.5.626}
\begin{equation}\label{vso}
V_{so} =-\frac{W_0}{2}\left(\rho \nabla \cdot \vec{J} + \sum_{\tau=n,p} \rho_\tau \nabla \cdot \vec{J}_\tau \right),
\end{equation}
with $\rho$ and $\vec{J}$ being the nucleon number density and spin-current density, respectively, and the subscript $\tau$ representing the isospin index. Besides $W_0$, the other 9 parameters in the Skyrme interaction can be expressed analytically in terms of 9 macroscopic quantities~\cite{Chen:2010qx} including the slope parameter $L$ of the symmetry energy, which has so far been constrained within about $30 \sim 90$ MeV from various probes~\cite{LI2013276,RevModPhys.89.015007}. In order to explore the largest neutron-skin effect, we mainly focus on results from $L=90$ MeV in the present study, while the values of other 8 macroscopic quantities are set to be their empirical values as listed in Table I of Ref.~\cite{Chen:2010qx}. The effects of $\lambda$th-order deformation $\beta_\lambda$ are included in the SHFB calculation by using the cylindrical transformed deformed harmonic oscillator basis~\cite{Stoitsov:2012ri}. For a given axial multipole moment associated with the deformation $\beta_\lambda$~\cite{Gambhir:1990uyn,Wang:2021tjg}, this code allows us to calculate the corresponding density distribution by the linear constraint method based on the approximation of the random phase approximation matrix. A more consistent calculation requires to achieve the ground state and the corresponding $\beta_\lambda$ with proper nuclear interaction parameters, which goes beyond the present scope. In the present study, we vary the deformed neutron skin by changing the value of $W_0$ under the constrained deformation parameters for $^{96}$Zr, $^{96}$Ru, $^{238}$U, and $^{197}$Au. Figure~\ref{fig1} displays the resulting nucleon density distributions in the $r_\perp-z$ plane, with $z$ representing the orientation of the symmetric axis and $r_\perp$ being perpendicular to $z$. We note that the definition of central tip-tip (body-body) collisions is the configuration with the $z$ ($r_\perp$) axis head-on. Since we have only constrained $\beta_2$ in $^{96}$Ru and $^{238}$U, components of $\beta_4$ seem to appear from the self-consistent SHFB calculation. While increasing $W_0$ slightly increases the RMS radius of the nuclei considered here, we find that $W_0$ generally has minor effects on the overall nucleon density distribution, since we have constrained $\beta_2$ and $\beta_3$ in the SHFB calculation.

\begin{figure}[!h]
\includegraphics[width=1\linewidth]{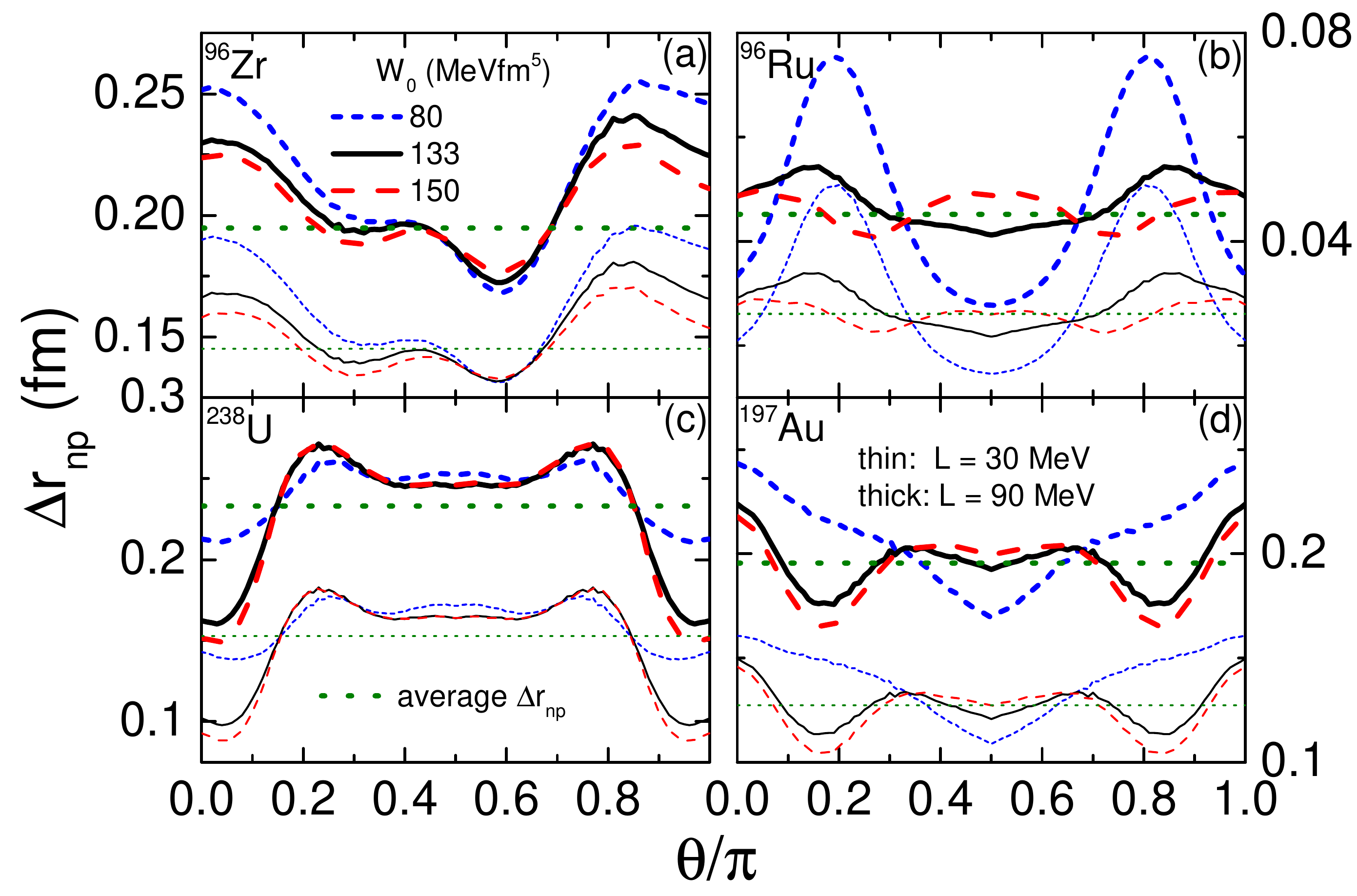}
\vspace*{-.3cm}
\caption{\label{fig2} Polar angular distribution of the neutron-skin thickness $\Delta r_{\mathrm{np}}$ in $^{96}$Zr (a), $^{96}$Ru (b), $^{238}$U (c), and $^{197}$Au (d) from deformed SHFB calculations using different slope parameters $L$ of the symmetry energy and different spin-orbit coupling constants $W_0$.}
\end{figure}

In deformed nuclei, the neutron-skin thickness $\Delta r_{\mathrm{np}}$ generally depends on the solid angle $\Omega=(\theta,\phi)$, i.e.,
\begin{eqnarray}
\Delta r_{\mathrm{np}}(\Omega) &=& \sqrt{\lr{r_\mathrm{n}^2(\Omega)}}-\sqrt{\lr{r_\mathrm{p}^2(\Omega)}},
\end{eqnarray}
where
\begin{equation}
    \sqrt{\lr{r_\mathrm{\tau}^2(\Omega)}} = \left(\frac{\int \rho_{\tau}(r,\Omega) r^4 d{r}}{\int \rho_{\tau}(r,\Omega) r^2 d{r} }\right)^{1/2}
\end{equation}
is the RMS radius for nucleons with isospin index $\tau$ in the direction $\Omega$. In the case of axial symmetry, the solid angular distribution $\Delta r_{\mathrm{np}}(\Omega)$ degenerates to a polar angular distribution $\Delta r_{\mathrm{np}}(\theta)$, with $\theta$ being the polar angle with respective to the symmetric axis $z$. Figure~\ref{fig2} compares the polar angular distributions of $\Delta r_{\mathrm{np}}$ in $^{96}$Zr, $^{96}$Ru, $^{238}$U, and $^{197}$Au obtained from deformed SHFB calculations by using different slope parameters $L$ of the symmetry energy and different spin-orbit coupling constants $W_0$. Generally, the nuclear spin-orbit coupling affects the nucleons in open shells consisting of spin unsaturated states, and these nucleons contribute significantly to the angular dependence of the neutron skin. The results thus depend on the numbers of protons and neutrons, and on the detailed shell structure of the particular nucleus. The average $\Delta r_{\mathrm{np}}$ from integrating $\Delta r_{\mathrm{np}}(\theta)$ over the polar angle $\theta$, which are sensitive to $L$ but nearly independent of $W_0$, are also plotted for comparison. The overall $\Delta r_{\mathrm{np}}$ is smaller with a smaller $L$ but its polar angular distribution is not affected by $L$. This is because the symmetry potential characterized by $L$ affects all nucleons in closed and open shells, so it has a global effect on the neutron-skin thickness. One sees that the distribution is symmetric with respective to $\theta=\pi/2$ for $^{96}$Ru, $^{238}$U, and $^{197}$Au with $\beta_3=0$, but asymmetric for $^{96}$Zr with $\beta_3>0$. While the detailed effect of $W_0$ on the polar angular distribution of $\Delta r_{\mathrm{np}}(\theta)$ depends on the nucleus, the appreciable sensitivity is observed in all cases. For $^{96}$Zr, a smaller $W_0$ leads to larger $\Delta r_{\mathrm{np}}$ around $\theta \sim 0$ and $\pi$ but slightly smaller $\Delta r_{\mathrm{np}}$ around $\theta \sim \pi/2$. For $^{96}$Ru, however, a smaller $W_0$ leads to significantly larger $\Delta r_{\mathrm{np}}$ around $\theta \sim \pi/5$ and $4\pi/5$ but reduces the $\Delta r_{\mathrm{np}}$ around $\theta \sim 0$, $\pi$, and $\pi/2$. For $^{238}$U, a larger $W_0$ enhances the $\Delta r_{\mathrm{np}}$ around $\theta \sim \pi/5$ and $4\pi/5$ but reduces the $\Delta r_{\mathrm{np}}$ around $\theta \sim 0$, $\pi$, and $\pi/2$. For $^{197}$Au, a larger $W_0$ enhances the $\Delta r_{\mathrm{np}}$ around $\theta \sim \pi/2$ but reduces the $\Delta r_{\mathrm{np}}$ especially around $\theta \sim \pi/5$ and $4\pi/5$. We have further found that the angular distributions of the RMS radii of the nuclei considered here are different from $\Delta r_{\mathrm{np}}(\theta)$, and are rather insensitive to $W_0$ with the constrained deformation parameters, while the ratios of $\Delta r_{\mathrm{np}}$ to the RMS radii have qualitatively similar $\theta$ dependencies as $\Delta r_{\mathrm{np}}(\theta)$.

The coordinates of initial neutrons and protons in colliding nuclei are sampled according to the above density distributions, while nucleon momenta are sampled isotropically within the isospin-dependent Fermi sphere, with the Fermi momentum calculated according to the local density of neutrons or protons. By using the nucleon-nucleon inelastic cross section of 42 mb at $\sqrtsnn=200$ GeV and 193 MeV, a Monte-Carlo Glauber model~\cite{Miller:2007ri} is then used to simulate the nucleus-nucleus collisions, based on which the participant nucleons and spectator nucleons are identified. In the present study, we discuss only central tip-tip collisions, body-body collisions, and collisions with random orientations at impact parameter $\text{b}=0$. While it is experimentally challenging to select both collision geometry and centrality, this is still possible by using zero degree calorimeters coupled with event-shape engineering~\cite{Goldschmidt:2015kpa}. The dynamics of participant matter is totally neglected, while the spectator matter obtained from the Glauber model are further grouped into heavy clusters ($A \geq 4$) and free nucleons based on a minimum spanning tree algorithm, i.e., nucleons with their distance $\Delta r<\Delta r_{\mathrm{max}}$ and relative momentum $\Delta p<p_{\mathrm{max}}$ may form heavy clusters. The coalescence parameters $\Delta r_{\mathrm{max}}=3$~fm and $\Delta p_{\mathrm{max}}=300$ MeV/$c$ taken from Ref.~\cite{Li:1997rc} have been shown to give the best description of the experimental data of free spectator neutrons in ultracentral \auau\ collisions at $\sqrtsnn=130$ GeV~\cite{Liu:2022kvz}. For spectator nucleons that do not form heavy clusters ($A \geq 4$), they may coalesce into light clusters with $A \leq 3$, i.e., deuterons, tritons, and $^3$He, and this process is implemented based on a Wigner function approach~\cite{Chen:2003ava,Sun:2017ooe}. The total free spectator nucleons are composed of the remaining neutrons and protons that have not coalesced into light clusters, and those from the deexcitation of heavy clusters. The deexcitation of heavy clusters with $A \geq 4$ are handled by the GEMINI model~\cite{Charity:1988zz,Charity:2010wk}, which requires as inputs the angular momentum and the excitation energy of the cluster. The angular momentum of the cluster is calculated by summing those from all nucleons with respective to their center of mass. The energy of the cluster is calculated from a simplified SHF energy-density functional~\cite{Chen:2010qx}, with the neutron and proton phase-space information obtained from the test-particle method~\cite{Wong:1982zzb,Bertsch:1988ik}, and its excitation energy is then calculated by subtracting from the calculated cluster energy the ground-state energy taken from the mass table~\cite{Wang:2021xhn} or an improved liquid-drop model~\cite{Wang:2014qqa}. For more details of the analysis procedure, we refer the reader to Refs.~\cite{Liu:2022kvz,Liu:2022xlm}.

\begin{figure}[ht]
\includegraphics[width=1\linewidth]{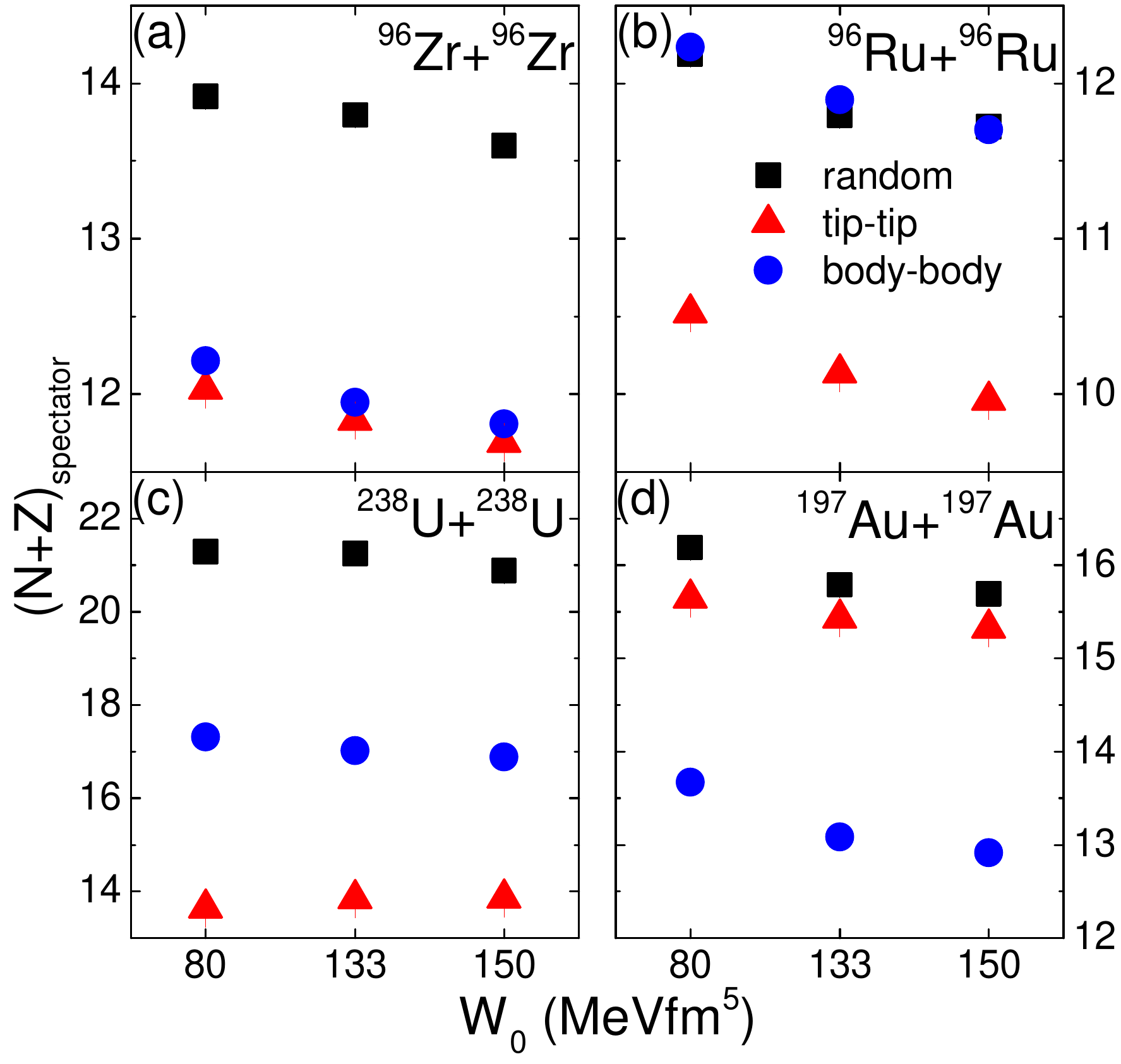}
\caption{\label{fig3} Total spectator nucleon numbers $N+Z$ in central \zrzr\ (a), \ruru\ (b), \uu\ (c), and \auau\ (d) collisions at top RHIC energy for different collision configurations and based on density distributions by using $L=90$ MeV and different $W_0$ in SHFB calculations.}
\end{figure}

\begin{figure}[!h]
\includegraphics[width=1\linewidth]{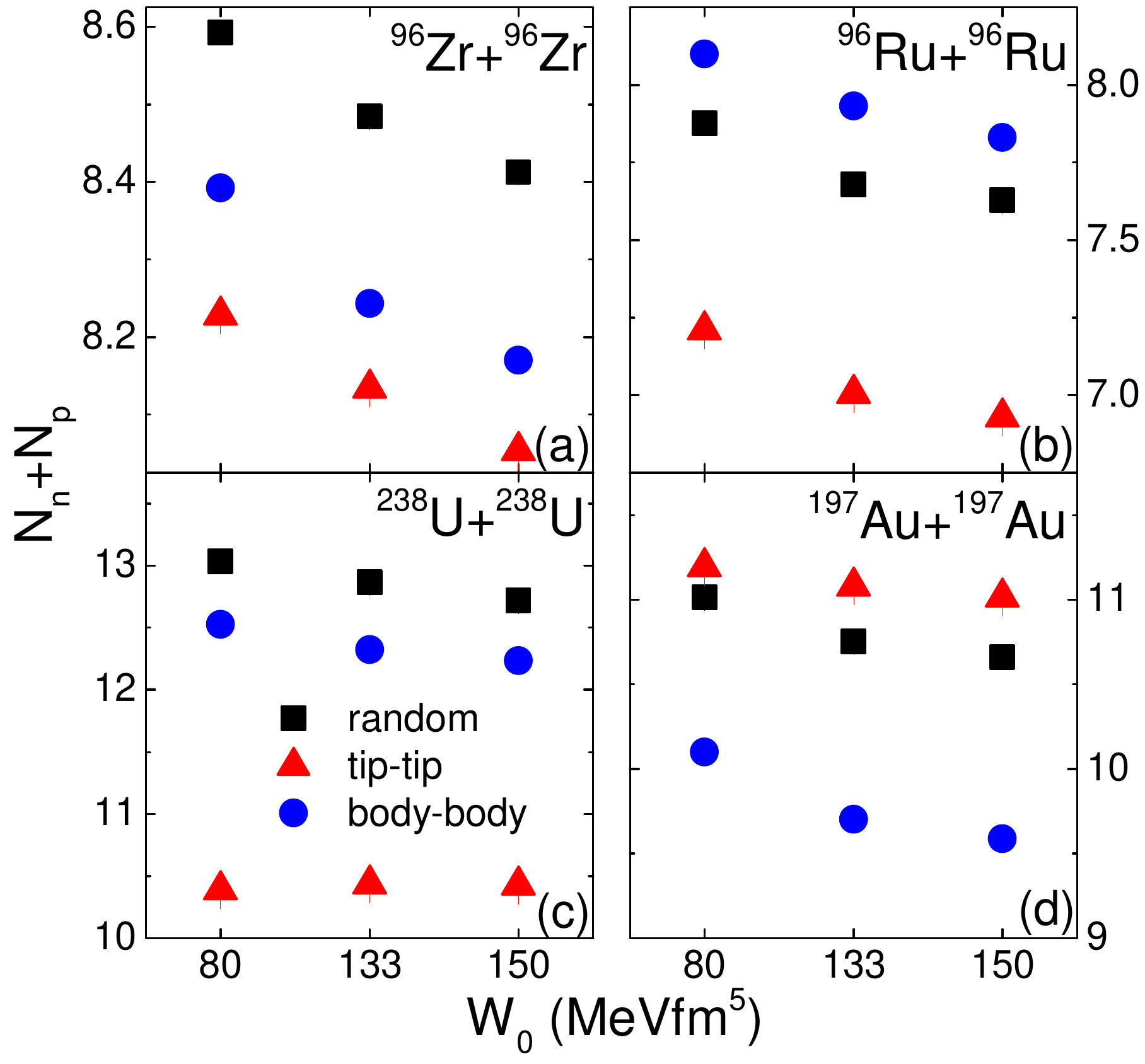}
\vspace*{-.3cm}
\caption{\label{fig4} Yield of free spectator nucleons $N_n+N_p$ in central \zrzr\ (a), \ruru\ (b), \uu\ (c), and \auau\ (d) collisions at top RHIC energy for different collision configurations and based on density distributions by using $L=90$ MeV and different $W_0$ in SHFB calculations. }
\end{figure}

The numbers $N+Z$ of total spectator nucleons in the four collision systems for different collision configurations and based on density distributions by using different $W_0$ are compared in Fig.~\ref{fig3}. Due to the slightly smaller RMS radii from larger $W_0$, $N+Z$ slightly decreases with increasing $W_0$ in all cases. In addition, collision systems with the colliding nuclei of a larger size or a larger neutron skin give an overall larger $N+Z$. One sees that collisions with random orientations generally lead to the maximum spectator nucleon number in all cases, while the values of $N+Z$ from tip-tip and body-body collisions depend on the constrained deformation parameters as well as their correlation with the detailed density distribution. For $^{96}$Zr with a small $\beta_2$ but a large $\beta_3$, tip-tip and body-body collisions lead to similar $N+Z$. For $^{96}$Ru and $^{238}$U with larger $\beta_2$, tip-tip \ruru\ and \uu\ collisions lead to the minimum spectator nucleon number compared to other collision configurations, while the value of $N+Z$ in body-body collisions depends on $\beta_2$. For $^{197}$Au with a negative $\beta_2$, body-body collisions lead to the minimum spectator nucleon number compared to other collision configurations. Free spectator nucleon numbers $N_n+N_p$ in the corresponding scenarios are compared in Fig.~\ref{fig4}. The dependencies on the collision configuration and the value of $W_0$ are similar to those in Fig.~\ref{fig3}, while the difference among different collision configurations becomes smaller compared with that for the total spectator nucleons, since about $30\%$ spectator nucleons are bound in clusters.

\begin{figure}[ht]
\includegraphics[width=1\linewidth]{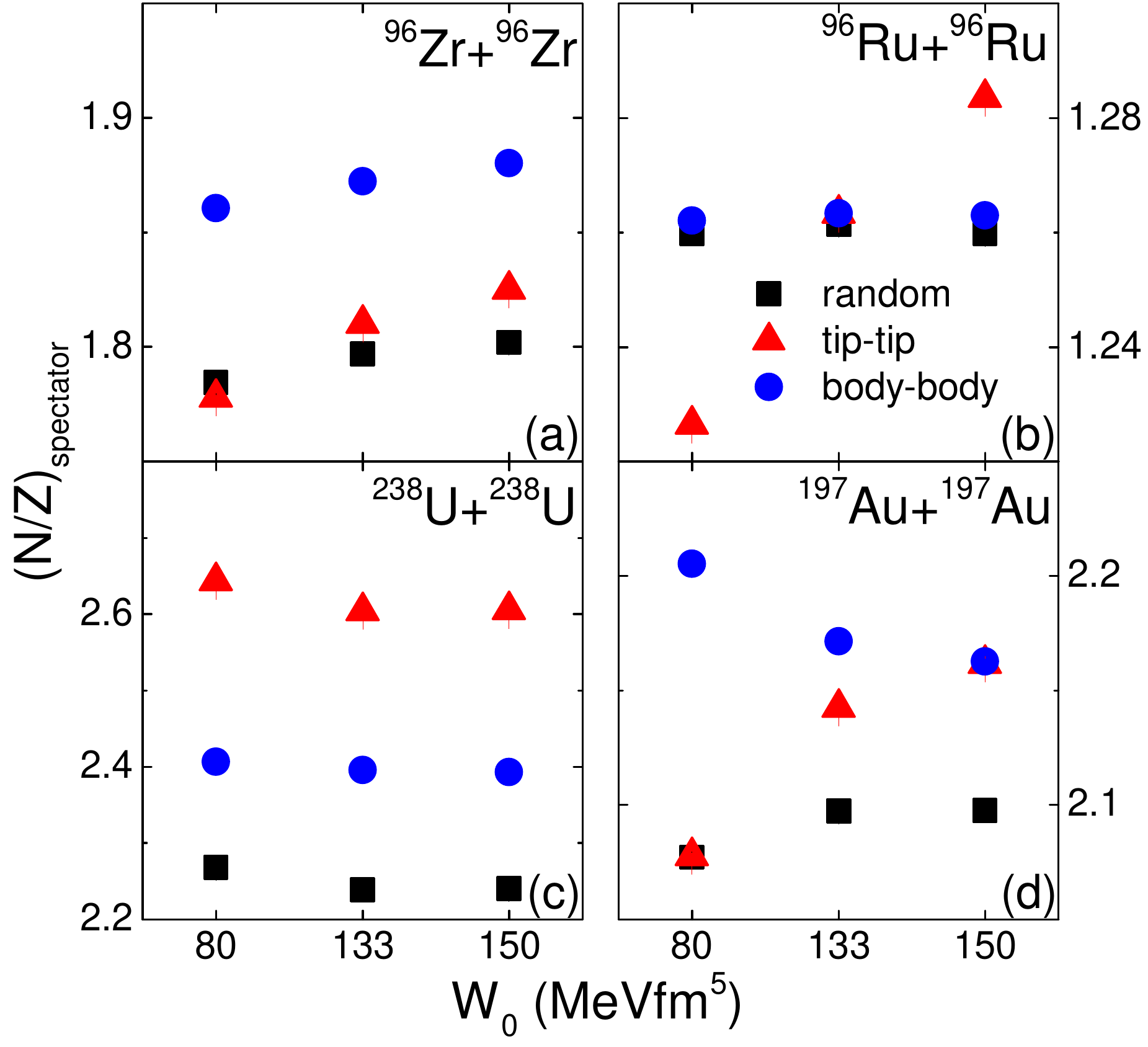}
\caption{\label{fig5} Ratios $N/Z$ of total spectator neutron to proton numbers in central \zrzr\ (a), \ruru\ (b), \uu\ (c), and \auau\ (d) collisions at top RHIC energy for different collision configurations and based on density distributions by using $L=90$ MeV and different $W_0$ in SHFB calculations.}
\end{figure}

\begin{figure}[!h]
\includegraphics[width=1\linewidth]{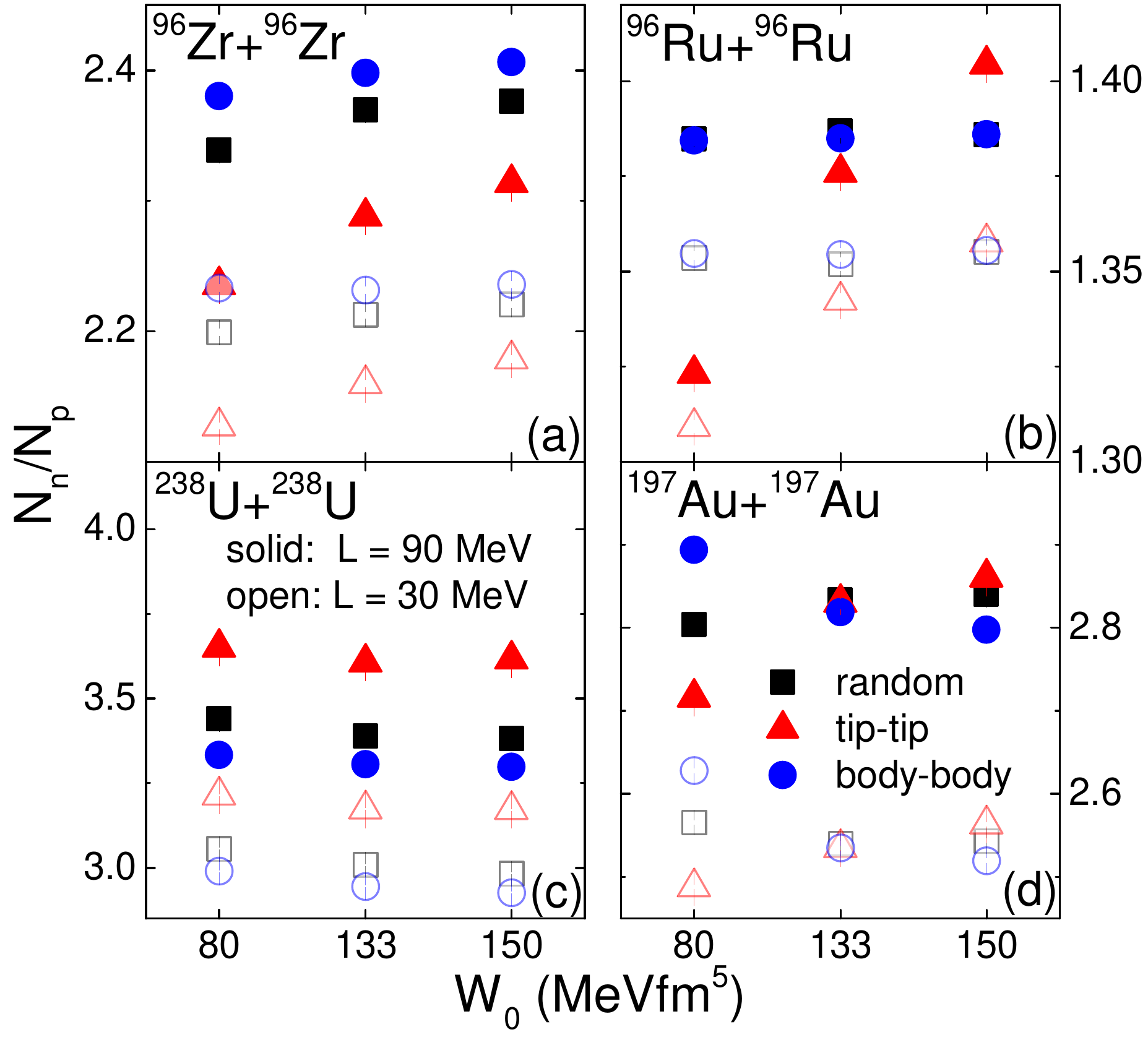}
\vspace*{-.3cm}
\caption{\label{fig6} Yield ratio $N_n/N_p$ of free spectator neutrons to protons in central \zrzr\ (a), \ruru\ (b), \uu\ (c), and \auau\ (d) collisions at top RHIC energy for different collision configurations and based on density distributions by using different $L$ and $W_0$ in SHFB calculations. }
\end{figure}

Figure~\ref{fig5} compares the ratios $N/Z$ of total spectator neutron to proton numbers in the four collision systems with different collision configurations and based on density distributions by using different $W_0$. The overall $N/Z$ is larger in a more neutron-rich collision system, where the colliding nuclei have larger neutron skins. For \zrzr\ collisions, the body-body collision configuration leads to the maximum $N/Z$ as a result of the larger $\Delta r_{\mathrm{np}}$ around $\theta \sim 0$ and $\pi$ compared to that around $\theta \sim \pi/2$. For \uu\ collisions, the tip-tip collision configuration leads to the maximum $N/Z$ as a result of the larger $\Delta r_{\mathrm{np}}$ around $\theta \sim \pi/2$ compared to that around $\theta \sim 0$ and $\pi$. While we haven't observed nontrivial dependence of the $N/Z$ value in \zrzr\ and \uu\ collisions for different collision configurations on $W_0$, results in \ruru\ and \auau\ collision systems need special attention. For \ruru\ collisions, the $N/Z$ value increases with increasing $W_0$ in the tip-tip collision configuration, different from that in body-body or random collision configuration. This is due to the increasing $\Delta r_{\mathrm{np}}$ around $\theta \sim \pi/2$ with increasing $W_0$, as seen from Fig.~\ref{fig2} (b). For \auau\ collisions, the $N/Z$ value increases with increasing $W_0$ in the tip-tip collision configuration but decreases with increasing $W_0$ in the body-body collision configuration. The former is due to the larger $\Delta r_{\mathrm{np}}$ around $\theta \sim \pi/2$ with a larger $W_0$, while the latter is due to the smaller $\Delta r_{\mathrm{np}}$ around $\theta \sim 0$ and $\pi$ with a larger $W_0$, as seen from Fig.~\ref{fig2} (d).

The isospin asymmetry of spectator matter discussed in Fig.~\ref{fig5} may manifest itself in the yield ratio $N_n/N_p$ of free spectator neutrons to protons, which can hopefully be measured experimentally, and the resulting $N_n/N_p$ ratios in different scenarios are compared in Fig.~\ref{fig6}, where results from both $L=90$ and 30 MeV are compared. Again, the qualitative dependencies of $N_n/N_p$ on the collision configuration and $W_0$ are similar to those of $N/Z$ as shown in Fig.~\ref{fig5}, while an overall larger $N_n/N_p$ than $N/Z$ is observed in all cases, since clusters are generally more isospin symmetric than free nucleons. By measuring the $N_n/N_p$ ratio in selected configurations of central \zrzr\ or \uu\ collisions, it is difficult to extract the value of $W_0$, since $N_n/N_p$ depends not only on $\Delta r_{\mathrm{np}}(\theta)$ but also on $L$. For \ruru\ and \auau\ collisions, however, one can extract the value of $W_0$ and thus probing the deformed neutron skin by simply comparing the $N_n/N_p$ ratio in central tip-tip and body-body collisions. For $L=90$ MeV, we found that the $N_n/N_p$ ratio is about $4\%$ ($7\%$) smaller in the tip-tip than body-body collision configuration in \ruru\ (\auau) collisions for $W_0=80$ MeVfm$^5$. For $L=30$ MeV, the overall $N_n/N_p$ ratio is smaller as already observed in Ref.~\cite{Liu:2022xlm}, while the difference between $N_n/N_p$ in central tip-tip and body-body \ruru\ collisions as well as in \auau\ collisions remains qualitatively similar although the magnitude becomes smaller for $W_0=80$ MeVfm$^5$. For $W_0=133$ MeVfm$^5$ or larger, the $N_n/N_p$ ratios are seen to be similar or slightly larger in central tip-tip \ruru\ and \auau\ collisions compared to those in the corresponding body-body collisions.

To summarize, we found that it is possible to measure the angular distribution of the neutron skin in deformed nuclei, by comparing the yield ratio $N_n/N_p$ of free spectator neutrons to protons in different collision configurations of central high-energy collisions with these nuclei. To illustrate the idea, we have obtained the neutron and proton density distributions of colliding nuclei consistently from the Skyrme-Hartree-Fock-Bogolyubov calculation, and varied the polar angular distribution of the neutron skin by adjusting the strength of the nuclear spin-orbit coupling. With the information of spectator nucleons obtained through a Monte-Carlo Glauber model, free spectator neutrons and protons are further generated through a multifragmentation process as detailed in Refs.~\cite{Liu:2022kvz,Liu:2022xlm}. By investigating the dependencies of the total and free spectator nucleon numbers as well as the corresponding neutron-proton asymmetries on the collision configuration and the deformed neutron skin in four typical collision systems, we found that although \zrzr\ and \uu\ collisions are not suitable for probing the deformed neutron skin by selecting special collision configurations, \ruru\ and \auau\ collisions are suitable systems for probing the angular distribution of the neutron skin by comparing the yield ratio $N_n/N_p$ in central tip-tip and body-body collisions. Because in $^{96}$Ru and $^{197}$Au, a weaker spin-orbit coupling leads to a smaller neutron skin around $\theta \sim \pi/2$ which results in a smaller $N_n/N_p$ ratio in central tip-tip collisions than in central body-body collisions, while a stronger spin-orbit coupling leads to a weaker polar angular dependence of the neutron skin and thus similar $N_n/N_p$ ratios in different collision configurations. To compare accurately the $N_n/N_p$ value in the two collision configurations, which is found to be qualitatively insensitive to the nuclear symmetry energy, requires a small systematic and statistical error from the experimental measurement. The proposed observables, if measured with dedicated zero-degree calorimeters in the corresponding heavy-ion experiments recently carried out by RHIC, may help to understand the structure of deformed nuclei as well as the interplay between the nuclear spin-orbit coupling and the symmetry energy.

We acknowledge helpful discussions with Jiangyong Jia, Chun-Jian Zhang, Xiang-Xiang Sun, and Shan-Gui Zhou. JX is supported by the National Natural Science Foundation of China under Grant No. 11922514. GXP and LML are supported by the National Natural Science Foundation of China under Grant Nos. 11875052, 11575190, and 11135011.

\bibliography{geometry}
\end{document}